\documentclass[12pt]{article}
\usepackage{epsf}
\def\be{\begin{equation}}
\def\ee{\end{equation}}
\def\bea{\begin{eqnarray}}
\def\eea{\end{eqnarray}}

\begin{document}
\begin{titlepage}
\begin{center}
{\Large \bf William I. Fine Theoretical Physics Institute \\
University of Minnesota \\}
\end{center}
\vspace{0.2in}
\begin{flushright}
FTPI-MINN-13/09 \\
UMN-TH-3141/13 \\
March 2013 \\
\end{flushright}
\vspace{0.3in}
\begin{center}
{\Large \bf Mixing of partial waves near $B^* \bar B^*$ threshold in $e^+e^-$ annihilation
\\}
\vspace{0.2in}
{\bf Xin Li$^a$  and M.B. Voloshin$^{a,b,c}$  \\ }
$^a$School of Physics and Astronomy, University of Minnesota, Minneapolis, MN 55455, USA \\
$^b$William I. Fine Theoretical Physics Institute, University of
Minnesota,\\ Minneapolis, MN 55455, USA \\
$^c$Institute of Theoretical and Experimental Physics, Moscow, 117218, Russia
\\[0.2in]

\end{center}

\vspace{0.2in}

\begin{abstract}
We consider the production of $B^* \bar B^*$ meson pairs in $e^+e^-$ annihilation near the threshold. The rescattering due to pion exchange between the mesons results in a mixing between three partial wave amplitudes: two $P$-wave with the total spin of the meson pair $S=0$ and $S=2$ and an $F$-wave amplitude. The mixing due to pion exchange with a low momentum transfer is calculable up to c.m. energy $E \approx 15 \div 20$\,MeV above the threshold. We find that the $P-F$ mixing is numerically quite small in this energy range, while the mixing of the two $P$-wave amplitudes is rapidly changing with energy and can reach of order one at such low energies. 
\end{abstract}
\end{titlepage}

The strong interaction between hadrons containing heavy $c$ or $b$ quarks as well as the light ones results in a peculiar  dynamics at energies where these hadrons move relatively slow with respect to each other. In particular,
the cross section for production of heavy meson pairs in $e^+e^-$ annihilation displays an intricate behavior near the thresholds for new meson pair channels. This behavior, observed for the onset of the production of charmed mesons~\cite{cleoc}as well as of the heavier $B$ and $B^*$ meson pairs~\cite{babar,bondar}, is not yet known in detail and further studies may uncover new hadronic structures at these thresholds and provide an insight into the dynamics of systems containing both heavy and light quark degrees of freedom. One important yet unknown feature of the production of pairs of heavy vector mesons is the spin-orbital structure of the production amplitudes. Namely, the overall quantum numbers $J^{PC}=1^{--}$ of a pair of e.g. $B^* \bar B^*$ mesons produced in the $e^+ e^-$ annihilation allow  for three different combinations of the orbital momentum and the total spin $S$ of the pair: $P$ wave with $S=0$ as well as a $P$ or $F$ wave with $S=2$. It is generally expected that the $F$ wave is kinematically suppressed near the threshold, which still leaves unknown the composition of the $P$-wave production amplitude in terms of the $S=0$ and $S=2$ components. The latter composition, clearly measurable from angular distributions~\cite{mv12}, can be quite nontrivial~\cite{dv} and in fact rapidly varying function of the c.m. energy in the near threshold region. In either case, the actual composition of the amplitude is very likely to be much different from a naive expectation~\cite{drgg,kmm,mv12} of dominance (by a factor 20 in the cross section) of the $S=2$ wave, following from the heavy quark spin symmetry, which symmetry is naturally broken in the threshold region~\cite{mv12}.

It is quite clear that the forces between the heavy mesons depending on the spins of light quarks result in a mixing of the partial waves, as recently discussed~\cite{mv13} for the $S -D$ mixing in the $J^P=1^+$ channel, and for heavy mesons the effect of these forces is enhanced by the factor of the meson mass $M$. There is likely a part of this interaction at short distances determined by $\Lambda_{QCD}$ that is currently impossible to analyze in a model independent way. However the interaction also contains a long-distance part due to the pion exchange (see e.g. in Ref.~\cite{mp}), determined by the strength $g$ of pion interaction with the heavy mesons, known~\cite{pdg} from the decay rate $D^* \to D \pi$. The effects of the pion exchange can be separated from those of the short-distance interaction in the range of c.m. energy above the threshold $E=p^2/M$  where the c.m. momentum $p$ is small compared to $\Lambda_{QCD}$, $p^2 \ll \Lambda_{QCD}^2$. Indeed, in the $J^{PC}=1^{--}$ channel two types of effects are possible: the $P-F$ wave mixing and the mixing of the $S=0$ and $S=2$ $P$ waves. The short-distance contribution to the $P-F$ wave mixing is determined by the parameter $M p^2/\Lambda_{QCD}^3$, while the pion contribution behaves~\cite{mv13} as $g^2 \, M/\Lambda_{QCD}$ at $p^2 \gg  \mu^2$ and as $g^2 \, M \, p^2/(\Lambda_{QCD} \mu^2)$ when $p$ is comparable with or smaller than the pion mass $\mu$. In what follows we calculate the effects arising from the pion exchange in the partial wave mixing for pairs of heavy vector $B^*$ mesons\footnote{A similar calculation is applicable also to the $D^* \bar D^*$ meson pairs. However the significance of the discussed effects at a fixed c.m. momentum is scaled down by the lighter mass of the charmed mesons.}, so that in our numerical estimates we use $M=5325\,$MeV. We find that the $P-F$ mixing, although parametrically enhanced at low $p$, is still of a moderate value and reaches only about 0.1 in the amplitude at the upper end of the applicability of our approach, $p \approx 300\,$MeV. Thus we confirm the existing expectation that the presence of the $F$ wave in production of heavy vector meson pairs in $e^+e^-$ annihilation is likely insignificant at energies slightly above the threshold.

The mixing of the $S=0$ and $S=2$ channels due to the short-distance interaction is generally of order one at any energy near the threshold. However, the energy scale for a variation of this part is set by $\Lambda_{QCD}^2/M$, so that no significant change is expected as long as $p^2 \ll \Lambda_{QCD}^2$. In particular, the absorptive part of the $S=0$ and $S=2$ mixing amplitude is proportional to $M \, p^3/\Lambda_{QCD}^4$ due to the $P$ wave phase space. On the other hand, the pion exchange contribution experiences a significant variation on a smaller energy scale. Namely we shall argue that the absorptive part of the $S=0$ and $S=2$ mixing amplitude behaves as $g^2 \, M \, p/\Lambda_{QCD}^2$ at $\mu^2 \ll p^2 \ll \Lambda_{QCD}^2$ and numerically changes from zero at the threshold to a factor of order one at $p \approx 200 \div 300\,$MeV. Thus the expected effect of the pion exchange in the latter mixing is a rapid variation above the threshold in the range of excitation energy up to $15 \div 20$\,MeV.

It can be also noticed that the re-scattering between the channels with two vector mesons and those with one or two pseudoscalar mesons, e.g. $B \bar B \to B^* \bar B^*$ and $B^* \bar B \to B^* \bar B^*$, which generally also contributes to the mixing of partial waves of the vector mesons, should be considered as a short-distance effect, whether it proceeds through the pion exchange or through other forces. Indeed, the momentum transfer in these processes is of the order of $q \sim \sqrt{M \, \Delta}$ with $\Delta$ being the mass difference between the vector and pseudoscalar mesons. Since in the heavy quark limit the parametric behavior is $\Delta \sim \Lambda_{QCD}^2/M$, one finds that in the cross-channel re-scattering $q \sim \Lambda_{QCD}$, and at such momentum transfer the pion exchange is indistinguishable from other short-range forces. For this reason in our calculations of the pion exchange  we consider only the diagonal re-scattering $B^* \bar B^* \to B^* \bar B^*$ as shown in Fig.~1

\begin{figure}[ht]
\begin{center}
 \leavevmode
    \epsfxsize=10cm
    \epsfbox{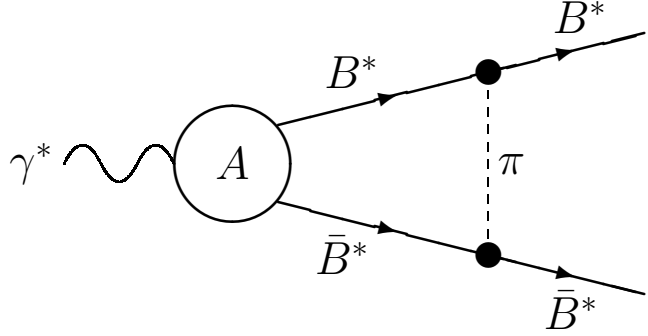}
    \caption{Diagonal re-scattering through pion exchange of $B^* \bar B^*$ meson pairs produced in $e^+e^-$ annihilation}
\end{center}
\end{figure}

The diagonal interaction of pions with the isotopic doublet of heavy vector mesons $V = (V^+, V^0)$ can be written as
\be
H_{int}=i \, {g \over f_\pi} \, \epsilon_{ljk} (V_j^\dagger \tau^a V_k) \, \partial_l \pi^a~,
\label{hint}
\ee
where $\tau^a$ are the isotopic Pauli matrices, $f_\pi \approx 132\,$MeV is the pion decay constant, and $g$ is a dimensionless coupling constant determined from the $D^* \to D \pi$ decay as $g^2 \approx 0.18$. It is assumed in Eq.(\ref{hint}) as well as throughout this paper that the nonrelativistic normalization is used for the wave functions of the heavy mesons.

The amplitude for the production of a $B^* \bar B^*$ pair in $e^+e^-$ annihilation near the threshold can generally be written in term of three partial wave amplitudes:
\bea
&&A(e^+e^- \to B^* \bar B^*) = A_0(p^2) \, j_k p_k \cdot {1 \over 3} \, a_l b_l + j_k \cdot {1 \over \sqrt{20}} \left ( a_i b_j +a_j b_i - {2 \over 3} \delta_{ij} \, a_l b_l \right ) \times \nonumber \\
&& \left \{ A_2(p^2) \,  \delta_{ki} p_j+ {5 \over \sqrt{6}} \, A_F(p^2) \, \left [ {1 \over p^2} \, p_i p_j p_k - {1 \over 5} \, \left (p_k \delta_{ij}+p_j \delta_{ik} + p_i \delta_{kj} \right ) \right ] \right \}~,
\label{3amp}
\eea
where $\vec p$ is the c.m. momentum of one of the mesons, $\vec a$ and $\vec b$ are the polarization amplitudes for the meson and anti-meson and $\vec j$ denotes the polarization amplitude of the virtual photon. The amplitudes $A_0$ and $A_2$ are the $S=0$ and $S=2$ $P$-wave amplitudes and $A_F$ is standing for the $F$-wave one. The relative normalization of the amplitudes in Eq.(\ref{3amp}) is chosen in such a way that the production cross section is proportional to $p^3 \, (|A_0|^2 + |A_2|^2 + |A_F|^2)$. One can also notice that under this normalization the expansion of $A_F$ at small momentum $p$ starts with $p^2$.

In what follows we treat the mixing of partial waves induced by the pion as a small effect and we calculate it in the first order of perturbation theory for which we use the nonrelativistic (in heavy mesons) formalism. Proceeding in this way and considering the projection on the $F$-wave we find the following expression for the amplitude $A_F$ generated after re-scattering through the pion exchange by the $P$-wave amplitudes $A_0$ and $A_2$:
\bea
&&A_F(p^2)= {g^2 \over f_\pi^2} \, {1 \over p^2} \, \left [ {1 \over p^2} \, p_i p_j p_k - {1 \over 5} \, \left (p_k \delta_{ij}+p_j \delta_{ik} + p_i \delta_{kj} \right ) \right ] \times \nonumber \\
&& \int {d^3 q \over (2 \pi)^3} \, \left [ {\sqrt{30} \over 2} \, A_0(q^2) + A_2(q^2) \right ] \, {M \over q^2-p^2 - i \epsilon} \, {q_i \, (q_j-p_j) \, (q_k-p_k) \over (\vec q - \vec p)^2+\mu^2} ~.
\label{afint}
\eea
One can notice that this expression includes the isotopic factor of 3, which corresponds to the pion exchange interaction in the isoscalar state of the $B^* \bar B^*$ pairs produced in $e^+e^-$ annihilation. 
The presence of the $F$ wave projector in Eq.(\ref{afint}) implies that only the part of the integral proportional to $p_i p_j p_k$ contributes to the $P-F$ mixing. Namely, if one writes the general expression allowed by the symmetry for the integral
\bea
&& \int {d^3 q \over (2 \pi)^3} \, \left [ {\sqrt{30} \over 2} \, A_0(q^2) + A_2(q^2) \right ] \, {M \over q^2-p^2 - i \epsilon} \, {q_i \, (q_j-p_j) \, (q_k-p_k) \over (\vec q - \vec p)^2+\mu^2} = \nonumber \\
&& C_1(p^2) \, p_i \, \delta_{jk} + C_2(p^2) \, (p_j \, \delta_{ik}+ p_k \, \delta_{ij}) + C_3(p^2) \, p_i p_j p_k~,
\label{c3}
\eea
only the structure proportional to $C_3$ contributes to the expression (\ref{afint}) for the amplitude $A_F$:
\be
A_F(p^2) = {2 \over 5} \, {g^2 \over f_\pi^2} \, p^2 \, C_3(p^2)~.
\label{afc3}
\ee 

Clearly, if the amplitudes $A_0$ and $A_2$ are smooth functions as $q^2$ varies on the scale of $p^2$ or $\mu^2$ the integral for $C_3$ converges and is determined by the range of $q$ such that $q^2 \sim p^2$, or $q^2 \sim \mu^2$ if $p^2 < \mu^2$. In order to estimate the numerical significance of the $P-F$ mixing we approximate the amplitudes $A_0$ and $A_2$ by constants, in which case the integral is calculated analytically and the result reads as
\be
A_F(p^2) = r_{FP} (p) \, \left [ {\sqrt{30} \over 2} \, A_0 + A_2 \right ]
\label{defrfp}
\ee
with the mixing function $r_{FP} (p)$ given by
\bea
&&r_{FP} (p)= {g^2 \over 20 \pi} \, {M \, \mu \over f_\pi^2} \,  \left \{ {5+ 18 \, t + 8 \, t^2 \over 16 t^{5/2}} \, \arctan (2 \sqrt{t}) - {15 + 34 t \over 24 t^2}  \right. \nonumber \\ 
&& \left. + i \, \left [ {15 +24 \,t - 4 \, t^2 \over 24 \, t^{3/2}} - {5+ 18 \, t + 8 \, t^2 \over 32 \, t^{5/2}} \, \log (1+4 \, t) \right ] \right \}~,
\label{rfp}
\eea
where $t=p^2/\mu^2$. 

The plot for the function $r_{FP} (p)$ is shown in Fig.~2. One can see that the discussed $P-F$ wave mixing is quite small. Indeed, at the upper end of the applicability range of our calculation, at $p \approx 300\,$MeV, the mixing function is still less than 0.05, corresponding to only of order 0.1 mixing with the $S=0$ $P$-wave amplitude, and smaller for the $S=2$ $P$-wave amplitude. 

\begin{figure}[ht]
\begin{center}
 \leavevmode
    \epsfxsize=10cm
    \epsfbox{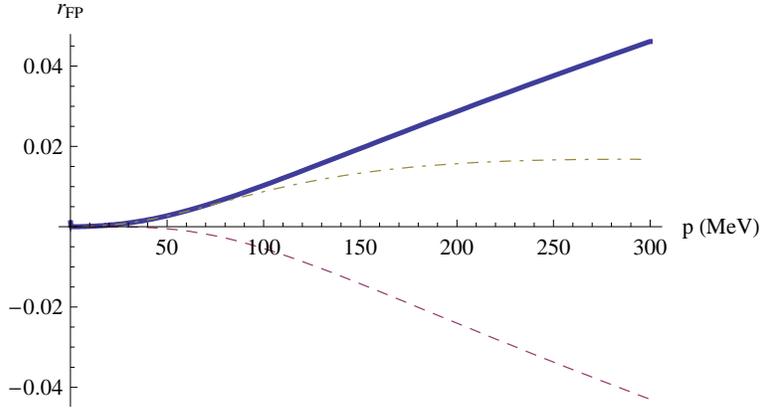}
    \caption{The function $r_{FP}$ describing the mixing of the $P$ and $F$ waves: the absolute value (solid), and the real (dotdashed) and imaginary (dashed) parts.}
\end{center}
\end{figure}

The available data~\cite{babar,bondar} do not indicate a presence of a resonance at the $B^* \bar B^*$ threshold in $e^+e^-$ annihilation. However the data are not yet conclusive and a threshold resonance may show up in future studies. In this case our approximation of smooth (constant) amplitudes $A_0$ and $A_2$ would generally be not applicable literally. For this reason we have verified that using in Eq.(\ref{afint}) these amplitudes with a Breit-Wigner shape instead of constants does not qualitatively change the conclusion that the $P-F$ mixing is small and remains at the level of 0.1 or less for a broad range of the resonance parameters.

We shall argue that the effect of the pion exchange is significantly larger numerically for the mixing of the two $P$-wave amplitudes $A_0$ and $A_2$. Proceeding to a calculation of this effect we notice that the mixing $A_0 \to A_2$ is the same as $A_2 \to A_0$ by reversibility, so that it is sufficient to consider the mixing only `in one direction', e.g. the $S=2$ amplitude $A_2$ generated by the $S=0$ production amplitude $A_0$. Using Eq.(\ref{hint}) for the heavy meson - pion interaction and our definition in Eq.(\ref{3amp}) of the production amplitudes, we find that for a pair initially produced by the amplitude $A_0$ an $S=2$ state is generated by rescattering with the amplitude
\bea
&&A_{\pi2}= j_k \, {1 \over \sqrt{20}} \, \left ( a_i b_j +a_j b_i - {2 \over 3} \delta_{ij} \, a_l b_l \right ) \times \nonumber \\
&&{\sqrt{5} \, g^2 \over f_\pi^2} \, \int {d^3 q \over (2 \pi)^3} \, A_0(q^2) \, {M \over q^2-p^2-i \epsilon} \, {q_k \, (q_i-p_i) \, (q_j-p_j) \over (\vec q- \vec p)^2 + \mu^2}~.
\label{api2}
\eea
If one writes the general expression for the three-index integral in terms of partial waves, 
\bea
&&\int {d^3 q \over (2 \pi)^3} \, A_0(q^2) \, {M \over q^2-p^2-i \epsilon} \, {q_k \, (q_i-p_i) \, (q_j-p_j) \over (\vec q- \vec p)^2 + \mu^2} = D_1(p^2) \, p_k \, \delta_{ij} \nonumber \\
&& + D_2(p^2) \, (p_j \, \delta_{ik} + p_i \, \delta_{jk}) + D_3(p^2) \, \left [ {1 \over p^2} \, p_i p_j p_k - {1 \over 5} \, \left (p_k \delta_{ij}+p_j \delta_{ik} + p_i \delta_{kj} \right ) \right ]~,
\label{d3}
\eea
it can be readily seen that the $S=2$ projector in Eq.(\ref{api2}) selects only the part proportional to the invariant function $D_2$. As a result the expression for the generated $S=2$ amplitude $\delta A_2$ can be written as
\be
\delta A_2(p^2)= {g^2 \over \sqrt{5} \, f_\pi^2} \, \int {d^3 q \over (2 \pi)^3} \, A_0(q^2) \, {M \over q^2-p^2-i \epsilon} \, {1 \over p^2} \, {3 \, (\vec p \cdot \vec q - p^2) \, (q^2- \vec p \cdot \vec q) - (\vec p \cdot \vec q) \, (\vec q- \vec p)^2 \over (\vec q- \vec p)^2 + \mu^2}~.
\label{da2}
\ee
Unlike the integral for the $P-F$ mixing in Eq.(\ref{afint}) this expression does not converge for a constant $A_0(q^2)$ and thus is not determined by the intermediate momentum $q^2 \sim p^2 \ll \Lambda_{QCD}^2$ if the amplitude $A_0(q^2)$ varies at the scale of $\Lambda_{QCD}^2$. Thus the pion exchange at small momentum transfer does not dominate the discussed mixing of the $P$ waves, and one should take into account other interactions at short distances. However the significance of the mixing generated by the pion exchange at longer distances can still be evaluated from Eq.(\ref{da2}) by considering the absorptive part of the mixing determined by $q^2=p^2$. The calculation of the absorptive part of the mixing amplitude is done by replacing in Eq.(\ref{da2}) the propagator $(q^2 - p^2- i \epsilon)^{-1}$ with $\pi \, \delta(q^2-p^2)$, and one finds
\be
\left. \delta A_2 (p^2) \right |_{\rm abs} = r_{20} (p) \, A_0(p^2)
\label{da2abs}
\ee
with the absorptive mixing function $r_{20}$ given by
\be
r_{20}(p)= {g^2 \over 16 \, \sqrt{5} \, \pi} \, { M \, p \over f_\pi^2} \left [ -6 + {\mu^2 \over p^2} + \left ( {\mu^2 \over p^2} - {\mu^4 \over 4 \, p^4 } \right ) \, \log \left ( 1 + {4 \, p^2 \over \mu^2} \right ) \right ]~.
\label{r20}
\ee

The plot of the function $r_{20}$ is shown in Fig.~3. One can see that this function changes between zero at the threshold to rather large values of about 0.75 at the upper end of the applicability of our calculation. This significant and rapid variation of the mixing in fact justifies a consideration of the absorptive part alone, since the dispersive part and any other effects arising from short distances are expected to exhibit a variation on the momentum scale of order $\Lambda_{QCD}$, which scale is parametrically larger than the range of the plot in Fig.~3. Thus any possible cancellation in the absorptive part due to the short-distance processes cannot take place in the entire range of momenta, at which our approach is applicable.  

\begin{figure}[ht]
\begin{center}
 \leavevmode
    \epsfxsize=10cm
    \epsfbox{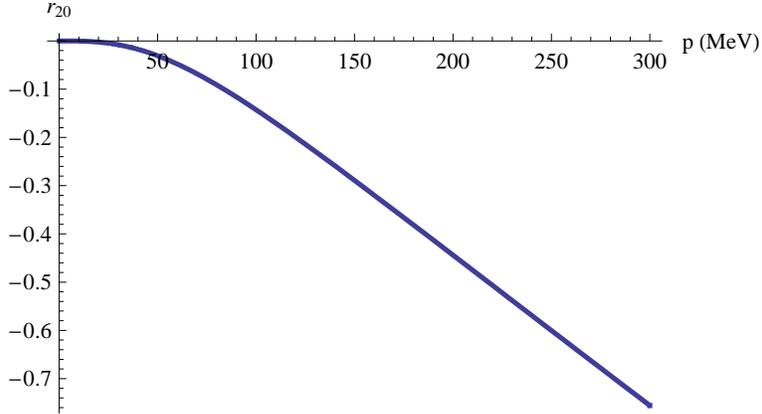}
    \caption{The function $r_{20}$ describing the absorptive part of the mixing between the $S=0$ and $S=2$ $P$-waves.}
\end{center}
\end{figure}

In summary. We have considered the effect of the pion exchange on the mixing between three partial waves of $B^* \bar B^*$ mesons produced in $e^+e^-$ annihilation at energy near the threshold. The pion exchange is calculable and dominates the mixing of $P$ and $F$ waves as long as the c.m. momentum $p$ of the mesons is small as compared to $\Lambda_{QCD}$, which restricts the range of the excitation energy of the meson pair to at most $E \approx 15 \div 20\,$MeV. We find that the $P-F$ mixing is rather small and should not exceed approximately 0.1 in the amplitude. The mixing effect is however significantly larger numerically for the mixing of the two $P$-wave states, corresponding to the total spin of the meson pair $S=0$ and $S=2$. Only for the absorptive part of this mixing the dominance of the pion exchange can be ensured, while the full effect generally depends on the unknown interaction at short distances determined by $\Lambda_{QCD}$. We find that in the latter case the absorptive part of the mixing rapidly changes with energy from zero at the threshold to about 0.75 at the upper end of the range where our approach is applicable. We thus conclude that the partial wave composition of the produced pairs of vector $B^*$ mesons should exhibit a nontrivial behavior near the threshold in $e^+e^-$ annihilation, which can be studied experimentally. 

The work of MBV is supported, in part, by the DOE grant DE-FG02-94ER40823.


\begin{thebibliography}{99}
\bibitem{cleoc} 
  R.~Poling,
  eConf C {\bf 060409}, 005 (2006)
  [hep-ex/0606016].
\bibitem{babar} 
  B.~Aubert {\it et al.}  [BABAR Collaboration],
  Phys.\ Rev.\ Lett.\  {\bf 102}, 012001 (2009)
  [arXiv:0809.4120 [hep-ex]].
\bibitem{bondar}
  A.~E.~Bondar [on behalf of the Belle Collaboration], Talk at the 36th Int. Conf. on High Energy Physics (ICHEP2012), Melbourn, July 2012. URL: http://belle.kek.jp/belle/talks/ICHEP12/A.Bondar.pdf
	
\bibitem{mv12} 
  M.~B.~Voloshin,
  Phys.\ Rev.\ D {\bf 85}, 034024 (2012)
  [arXiv:1201.1222 [hep-ph]].
	
\bibitem{dv} 
  S.~Dubynskiy and M.~B.~Voloshin,
  Mod.\ Phys.\ Lett.\ A {\bf 21}, 2779 (2006)
  [hep-ph/0608179].	
	
\bibitem{drgg} 
  A.~De Rujula, H.~Georgi and S.~L.~Glashow,
  Phys.\ Rev.\ Lett.\  {\bf 38}, 317 (1977).
	
\bibitem{kmm} 
  R.~Kaiser, A.~V.~Manohar and T.~Mehen,
  Phys.\ Rev.\ Lett.\  {\bf 90}, 142001 (2003)
  [hep-ph/0208194].
	
\bibitem{mv13} 
  M.~B.~Voloshin,
  arXiv:1301.5068 [hep-ph].

\bibitem{mp}
  T.~Mehen and J.~W.~Powell,
  Phys.\ Rev.\ D {\bf 84}, 114013 (2011)
  [arXiv:1109.3479 [hep-ph]].
	
\bibitem{pdg}
J. Beringer et al. (Particle Data Group), Phys.\ Rev.\ D {\bf 86}, 010001 (2012) 

\end{thebibliography}
\end{document}